\documentclass[11pt]{JHEP3}
\usepackage{amsfonts}
 \usepackage{amssymb}
 \usepackage{amsmath}
 \newfont{\bbbold}{msbm10}

 \def\cL{{\cal L}}

 \newfont{\goth}{eufm10 scaled \magstep1}

 \def\a{\alpha}
 \def\b{\beta}
 \def\c{\gamma}
 \def\d{\delta}
 \def\e{\epsilon}

 \def\m{\mu}
 \def\n{\nu}

 \def\t{\tau}

 \def\be{\begin{equation}}\def\ee{\end{equation}}
 \def\bea{\begin{eqnarray}}\def\eea{\end{eqnarray}}
 \def\ba{\begin{array}}\def\ea{\end{array}}

 \def\del{\partial}

 \def\str{\rm str}

 \def\del{\partial}

 \def\3dt{\dot{3}}

 {}

 \def\bd{\begin{document}}
 \def\ed{\end{document}}
 \def\bea{\begin{eqnarray}}
 \def\ba{\begin{array}}\def\ea{\end{array}}
 \def\eea{\end{eqnarray}}
 \def\ft#1#2{{\textstyle{{\scriptstyle #1}\over {\scriptstyle #2}}}}
 \def\fft#1#2{{#1 \over #2}}
 \newcommand{\eq}[1]{(\ref{#1})}
 \def\eqs#1#2{(\ref{#1}-\ref{#2})}
 \def\det{{\rm det\,}}
 \def\tr{{\rm tr}}\def\Tr{{\rm Tr}}
  \def\str{{\rm str}} \def\diag{{\rm diag}}
 \def\sdet{{\rm sdet}}\def\symtr{{\rm symtr}}

\newcommand{\hoch}[1]{$^{#1}$}

\title{Universal Subleading Spectrum of Effective String Theory}

\author{J. M. Drummond\\ Department of Mathematics, Trinity College
  Dublin\\ E-mail: \email{jmd@maths.tcd.ie}}

\abstract{We analyse the spectrum of the D-dimensional Poincar\'e
  invariant effective 
string model of Polchinski and Strominger. It is shown that the leading terms
beyond the Casimir term in the long distance expansion of the spectrum have a
universal character which follows from the constraint of Poincar\'e
  invariance.} 

\begin{document}

%%%%%%%%%%%%%%%%%%%%%%%%%%%%%%%%%%%%%%%%%%%%%%%%%%%%%%%%%%%%%%%%%
%%%%%%%%%%%%%%%%%%%%%%%%%%%%%%%%%%%%%%%%%%%%%%%%%%%%%%%%%%%%%%%
\section{Introduction}
The question of whether QCD can be described by a string theory is one which
has commanded much attention over several decades but is still unresolved. It
is widely believed that the colour flux between a quark-antiquark pair can be
described as a string for sufficiently large separations. The question of
which, if any, string theory is relevant for such systems is a difficult
problem. In \cite{ps91} Polchinski and Strominger showed that it was possible
to construct a string theory with manifest D-dimensional Poincar\'e
symmetry. This string theory evades the traditional restriction to the
critical dimension of 26 by including a term in the action which is
valid only for expansion around a `long string' vacuum, i.e. it diverges if
the string is allowed to shrink to zero size. The coefficient of the new
term can then be adjusted to cancel the anomalous central charge outside 26
dimensions. 

We discuss the effective string model of \cite{ps91} and show
that the constraints of D-dimensional Poincar\'e invariance imply a universal
subleading behaviour in the spectrum. Specifically we show that the
$R^{-3}$ terms are fixed and have the same form as in the Nambu-Goto spectrum.
Such higher order terms in the spectrum are relevant
for comparisons to lattice simulations of QCD flux tubes and other string-like
solitons \cite{lw02,jkm03,jkmmp03,jkm04}.

\section{Effective String Model}

In this section we very briefly reexamine the D-dimensional covariant 
string model \cite{ps91} and we refer the reader to the original article for
further detail.   
The effective string action given in \cite{ps91} is 
\be
S = \frac{1}{4 \pi} \int d \t^+ d \t^-  
\left[ 
\frac{1}{a^2} \del_+ X \cdot \del_- X 
+ \b \frac{\del_+^2 X\cdot\del_- X \del_+ X\cdot\del_-^2 X}
          {(\del_+ X\cdot\del_- X)^2}           
\right].
\ee
As noted in \cite{ps91} the non-polynomial terms in the action are no problem
as long as one only considers expanding the theory about a `long string'
vacuum where the first derivatives $\del_+ X$ and $\del_- X$ are never
small. Accordingly we consider the string to be wrapped around a spatial
dimension compactified on a circle of radius $R$ \footnote{Such an object is
  referred to as a `torelon' in the QCD flux tube context.}. The ground state
is given by  
\be
X^{\m}_{\rm cl} = e_+^\m R \t^+ + e_-^\m R \t^-.
\ee  
The first derivatives of $X$ are then of order $R$ (which we take to be large)
and higher derivatives are order one.

Adopting the notation that $Z = \del_+ X\cdot\del_- X$, we write this as
\be
S = \frac{1}{4 \pi} \int d\t^+ d\t^-
\left[ 
\frac{1}{a^2} Z  + 
\b \frac{\del_+^2 X \cdot \del_- X \del_+ X\cdot\del_-^2 X}{Z^2}
\right].
\ee
We will call the two terms $S_0$ and $S_\b$ respectively. 

The new term in the action was motivated by the fact that in a properly
covariant treatment of the underlying field theory one would expect some
determinants to appear on the transformation from the field theory variables to
the string theory variables. The new term is then precisely the Polyakov
determinant in terms of the induced metric instead of an intrinsic metric.
One can check that the term $S_\b$ is the only term up to $O(R^{-2})$ which
obeys the requirements that it must have inverse powers only of the operator
$Z=\del_+ X \cdot \del_- X$ (which has a large classical expectation value) and
is not proportional to the lowest order equations of motion $\del_+ \del_- X
=0$ (otherwise it would be removable by a field redefinition) or the lowest
order energy-momentum tensor $\del_- X \cdot \del_- X$ or $\del_+ X \cdot
\del_+ X$ (which vanishes between physical states). 

It is worth noting that one can 
rewrite the term $S_\b$ in a slightly simpler form. Using integration by parts
\footnote{We have no boundary and although we wrap the string around a compact
  dimension, the fields $\del_+ X$ and $\del_- X$ are still periodic.} one
  finds 
\be
S = \frac{1}{4 \pi} \int d\t^+ d\t^-
\left[ 
\frac{1}{a^2} Z  + 
\b \frac{\del_+^2 X \cdot \del_-^2 X}{Z}
\right].
\ee
The modified conformal transformation law,
\be
\d X = \e^-(\t^-) \del_- X - \frac{\b a^2}{2} \del_-^2 \e^-(\t^-) \frac{\del_+
  X}{Z} 
+  (+ \longleftrightarrow -), 
\ee
can be written as $\d X = \d_0 X + \d_\b X$ and is such that
\be
\d_0 S_0 = 0 , \hspace{20pt} \d_0 S_\b + \d_\b S_0 = 0.
\ee
These equations holds exactly, i.e. to all orders in $R^{-1}$, a point not
stressed in the original paper \cite{ps91}. 
The non vanishing term $\d_\b S_\b$ must be cancelled by appropriate
extra terms, $S_{\b^2}$ in the Lagrangian and $\d_{\b^2} X$ in the
transformation law. Thus the perturbation $S_\b$ generates an infinite series
of extra terms to make the Lagrangian invariant. The continuation of the
invariant at higher orders is not necessarily unique and the presence of free
parameters signals the start of a new invariant.

The term $\d_\b S_\b$ is $O(R^{-4})$ in
the expansion of inverse string length. This implies that the required
corrections to the action and transformation law only produce variations at
this order and 
therefore corrections to the energy momentum tensor at $O(R^{-3})$.
Any other new term $S_\c$ in the action must obey the same constraints which
were applied in the construction of $S_\b$, namely that it must have inverse
powers only of the operator $Z=\del_+ X \cdot \del_- X$ and that it must not be
proportional the lowest order equations of motion or the lowest order energy
momentum tensor. The first new
terms which satisfy these requirements are in fact of order $R^{-6}$
\footnote{except for pseudoscalar terms, e.g. a term in D=4 which is of order
  $R^{-2}$.} and a basis which makes this explicit is:  
\begin{align}
L_1 &= \frac{1}{Z^3} \del_-^2 X \cdot \del_-^2 X \del_+^2 X \cdot
\del_+^2 X, \label{L1}
\\ 
L_2 &= \frac{1}{Z^3} \del_-^2 X \cdot \del_+^2 X \del_-^2 X \cdot
\del_+^2 X, \label{L2}
\\ 
L_3 &= \frac{1}{Z^4} \del_-^2 X \cdot \del_+^2 X \del_- X \cdot
\del_+^2 X \del_-^2 X \cdot \del_+ X, \label{L3} \\
L_4 &= \frac{1}{Z^5} \del_- X \cdot \del_+^2 X \del_- X \cdot \del_+^2 X
\del_-^2 X 
\cdot \del_+ X \del_-^2 X \cdot \del_+ X. \label{L4}
\end{align}
The coefficients of these terms will be constrained by the requirements of
classical and quantum conformal invariance.
Given all of the above we certainly expect the energy momentum tensor derived
just from the the first corrections, $S_\b$ and $\d_\b X$, to be valid up to
$O(R^{-2})$ and we will show that it is rather simple to deduce the physical
spectrum up to $O(R^{-3})$ from these terms. 

From the lowest order action and transformation law we find the standard
energy-momentum tensor at lowest order,
\be
T^{0}_{--} = -\frac{1}{2a^2} \del_- X\cdot \del_- X,
\ee
The order $\b$ corrections to the action and transformation law give (after
some calculation), 
\begin{align}
T^{\b}_{--} = -\frac{\b}{2} 
\Bigl[
&-\frac{1}{Z}\del_-^3 X\cdot \del_+ X
+\frac{1}{Z^2}(-\del_- X\cdot \del_- X \del_-^2 X\cdot \del_+^2 X 
\notag \\
&+\del_-^2 X\cdot \del_- X \del_- X\cdot \del_+^2 X
               +\del_-^2 X\cdot \del_+ X \del_-^2 X\cdot \del_+ X)
\Bigr]. 
\end{align}
To verify that this is conserved, one must calculate the equation of
motion for $X$. Up to $O(R^{-3})$ this reads,
\begin{align}
\del_+ \del_- X = &\frac{\b a^2}{2 Z^2} 
(\del_- X \del_-^2 X\cdot \del_+^3 X + \del_+ X \del_-^3 X\cdot\del_+^2 X
-\del_-^2 X \del_- X\cdot \del_+^3 X - \del_+^2 X \del_-^3 X\cdot\del_+ X  )
\notag \\
&+ O(R^{-4}) .\label{eomXform}
\end{align}
Employing this one finds that $\del_+ T_{--} = O(R^{-3})$. We now expand this
in terms of the fluctuation field, $Y$, which is defined by
\be
X^\m = e_+^\m R \t^+ + e_-^\m R \t^- + Y^\m. \label{Yfield}
\ee
The lowest order constraints and periodicity imply
\be
e_+ \cdot e_+ = e_- \cdot e_- = 0 \text{ and } e_+ \cdot e_- = -\frac{1}{2}.
\ee
We then have 
\begin{align}
T_{--} = &-\frac{R}{a^2} e_-\cdot \del_- Y - \frac{1}{2 a^2} \del_- Y\cdot
\del_- Y 
-\frac{\b}{R} e_+\cdot \del_-^3 Y 
\notag \\ 
       &- \frac{\b}{R^2} \del_+ Y\cdot\del_-^3 Y
       -\frac{2 \b}{R^2} e_+\cdot \del_-^3 Y
        (e_+ \cdot \del_- Y + e_- \cdot \del_+ Y)
\notag \\
       &-\frac{2 \b}{R^2}(\del_+^2Y \cdot e_-\del_-^2Y\cdot e_- + e_+\cdot
\del_-^2 Y e_+ \cdot \del_-^2 Y) 
       +O(R^{-3}). \label{emtensorYform}
\end{align}

We can also expand the Lagrangian in terms of $Y$,
\begin{align}
\cL = &-\frac{R^2}{8\pi a^2} + \frac{1}{4\pi a^2} \del_+ Y \cdot \del_- Y
      +\frac{\b}{\pi R^2}\del_-^2 Y \cdot e_+ e_- \cdot \del_+^2 Y 
\notag \\
      &+\frac{\b}{\pi R^3} [\del_+^2 Y \cdot e_- \del_+ Y \cdot \del_-^2 Y
       +\del_+^2 Y \cdot \del_- Y e_- \cdot \del_-^2 Y]
\notag \\
      &+\frac{4 \b}{\pi R^3} \del_+^2 Y \cdot e_- e_+ \cdot \del_+^2 Y
       [e_+ \cdot \del_- Y + e_- \cdot \del_+ Y]  +
      O(R^{-4}). \label{lagrangianYform} 
\end{align}
In fact, after a field redefinition, this can be written as
\be
\cL = -\frac{R^2}{8\pi a^2} 
+ \frac{1}{4\pi a^2} \del_+ \hat{Y} \cdot \del_- \hat{Y} + O(R^{-4}),
\label{lagrangianYhatform}
\ee
where
\begin{align}
\hat Y &= Y - \frac{\b a^2}{R^2} 
[e_- \del_+ \del_- Y \cdot e_+ + e_+ \del_+ \del_- Y \cdot e_-]
\notag \\
&+ \frac{2\b a^2}{R^3}
[\del_- Y e_- \cdot \del_+^2 Y + \del_+ Y e_+ \cdot \del_-^2 Y
-\del_+ Y \del_+ \del_- Y \cdot e_- - \del_- Y \del_+ \del_- Y \cdot e_+
\notag \\
&-e_- \del_- Y \cdot \del_+^2 Y - e_+ \del_+ Y \cdot \del_-^2 Y
-4e_- e_+ \cdot \del_- \del_+ Y e_+ \cdot \del_- Y
-4e_+ e_- \cdot \del_+ \del_- Y e_- \cdot \del_+ Y]. \label{redef}
\end{align}
To see this, one must use partial integrations to rewrite the
correction terms in $\cL$ so that they are proportional to $\del_+ \del_-
Y$. This is the variation of the lowest order Lagrangian and so these
terms can be removed by field redefinition. This procedure cannot be continued
to higher orders as one can see from expanding the Lagrangian to the next
order in $R^{-1}$.
Since the Lagrangian is just the free field Lagrangian
for the new field, operator products for this field can be evaluated
just as in free field theory.  

Inverting (\ref{redef}) and substituting into (\ref{emtensorYform}),
the energy momentum tensor is found to be
\begin{align}
T_{--} = &-\frac{R}{a^2} e_-\cdot \del_- \hat Y 
        - \frac{1}{2 a^2} \del_- \hat Y \cdot \del_- \hat Y
       -\frac{\b}{R} e_+ \cdot \del_-^3 \hat Y 
\notag \\ 
       &-\frac{2 \b}{R^2} (e_+ \cdot \del_-^3 \hat Y 
         e_+ \cdot \del_- \hat Y 
        -e_+ \cdot \del_-^2 \hat Y 
         e_+ \cdot \del_-^2 \hat Y)
       +O(R^{-3}). \label{emtensorYhatform} 
\end{align}
One can then obtain the $TT$ operator product from the simple form of the
$\hat Y \hat Y$ operator product, 
\be 
T_{--}(\t^-)T_{--}(0) = \frac{12 \b + D}{2 (\t^-)^4} +
\frac{2}{(\t^-)^2} T_{--}(0) + \frac{1}{\t^-} \del_- T_{--}(0) +
O(R^{-2}),  
\ee
as required for conformal invariance. Note that we have this formula without
considering higher order corrections to the action or to the transformation
law (since their effect appears at higher order in $R^{-1}$). 
This formula implies the Virasoro algebra for the
corresponding conserved charges, $L_n$, 
\be
[L_m,L_n] = (m-n)L_{m+n} +\frac{12 \b + D}{12}(m^3 - m) \d_{m,-n},
\ee
where the central charge is $12 \b + D$ which we require to take the
value $26$ for a critical string. As noted in \cite{ps91} this fixes the value
of the parameter $\b$,
\be
\b = \frac{26 - D}{12}.
\ee 

The field $\del_- \hat Y$ obeys $\del_+ \del_- \hat Y = O(R^{-4})$ and hence
has an expansion of the form
\be
\del_- \hat Y^{\m} = a \sum_{m = -\infty}^{\infty} \hat \a_{m}^{\m}
e^{-im\t^-} + O(R^{-4}). 
\ee

The operators $\hat \a_{m}^{\m}$ satisfy the standard algebra up to
$O(R^{-4})$ since the field $\hat Y$ obeys the free field OPE up to
this order. It is then simple to express the Virasoro generators in
terms of the mode operators, $\hat \a$. We have, including a possible
normal ordering constant,
%\be
%T_{--}(\t^-) = \sum_{n=-\infty}^{\infty} L_n e^{-in\t^-}  
%\ee
%where
\begin{align}
L_n = &\frac{R}{a}e_- \cdot \hat \a_n 
+ \frac{1}{2}\sum_{m=-\infty}^{\infty}
:\hat \a_{n-m} \cdot \hat \a_{m}: 
+\frac{\b}{2}\d_{n,0} 
\notag \\
-&\frac{\b an^2}{R} e_+ \cdot \hat \a_{n}
-\frac{\b a^2 n^2}{R^2} e_{+ \m} e_{+ \n} 
\sum_{m=-\infty}^{\infty} \hat \a_{n-m}^{\m} \hat \a_{m}^{\n} 
+O(R^{-3}).
\end{align}
The $\d_{n,0}$ term can be calculated using the Virasoro algebra and
the standard algebra for the mode operators. This is the crucial
formula. It determines the form of the spectrum up to $O(R^{-3})$ as
we show below.

The spacetime momentum is given by
\be
p^{\m} = \frac{R}{2a^2}(e_-^{\m} + e_+^{\m}) 
+ \frac{1}{2a}(\a^{\m}_0 + \tilde{\a}^{\m}_0) + O(R^{-4}).
\ee

Imposing the physical condition $L_0 = \tilde{L}_0 = 1$ we find a universal
form for the $O(R^{-3})$ correction to the ground state energy,
\be
(-p^2)^{\frac{1}{2}} = \frac{R}{2a^2} + \frac{\b - 2}{R} - \frac{a^2}{R^3}
(\b - 2)^2 + O(R^{-4}),
\ee
with the parameter $\b$ fixed to be $(26-D)/12$.
We expect that free parameters will enter the spectrum at higher
orders. One
can go on and check explicitly that there are still only $(D-2)$ physical
oscillations at the first excited level, a fact guaranteed by the critical
value of the central charge.

\section{Conclusion}
We have shown that the Poincar\'e invariant effective string model predicts a
universal spectrum of fluctuations up to and including the $R^{-3}$
term where $R$ is the string length. This result was obtained without
the need to include terms at higher order in the Lagrangian because
they can only affect the spectrum at yet higher orders. 
We have carried out the analysis
for a wrapped closed string (or torelon) and have not included
possible boundary effects for open strings. It is interesting to note,
however, that such boundary effects are not expected to influence the
spectrum at the order we are considering \cite{lw04}.  
The spectrum obtained is precisely the spectrum of the Nambu-Goto
string up to and including the $R^{-3}$ term, with Poincar\'e
invariance being expected to force the two to differ at higher orders.

So far we have found no free parameters in the spectrum and hence the
terms obtained at $R^{-3}$ are {\sl universal}. The only assumption
that has been made in the derivation is that the effective string
model should be Poincar\'e invariant.
It would be very interesting to continue the analysis to higher orders to
quantify the number of free parameters appearing there. It is certainly worth
remarking that the terms in the Lagrangian required for classical conformal
invariance  actually appear at $O(R^{-6})$ thus one might expect the universal
behaviour in the spectrum to persist even beyond $R^{-3}$. 

\acknowledgments

The author would like to thank KJ. Juge, J. Kuti, F. Maresca,
C. Morningstar and M. Peardon for interesting discussions.

\end{document}